\documentclass{webofc}
\usepackage[varg]{txfonts}   
\woctitle{QCD@Work 2014}
\begin{document}
\title{Rapidity resummation for $B$-meson wave functions}
%
%

\author{
        Yue-Long Shen \inst{1}\fnsep\thanks{\email{shenylmeteor@ouc.edu.cn} } \and
        Yu-Ming Wang\inst{2,3}\fnsep\thanks{\email{yuming.wang@tum.de}}
}

\institute{
College of Information Science and Engineering,
Ocean University of China, Qingdao, Shandong 266100, P.R. China
\and
Physik Department T31, Technische Universit\"at M\"unchen,
James-Franck-Stra\ss e~1, D-85748 Garching, Germany
\and
Institut f\"ur Theoretische Teilchenphysik und Kosmologie,
RWTH Aachen University, D-52056 Aachen, Germany
}

\abstract{%
Transverse-momentum dependent (TMD) hadronic wave functions develop light-cone divergences
under QCD corrections, which are commonly regularized by the rapidity $\zeta$ of gauge vector defining
the non-light-like Wilson lines. The yielding rapidity logarithms from infrared enhancement need to be resummed
for both hadronic wave functions and short-distance functions, to achieve scheme-independent calculations
of physical quantities.  We briefly review the recent progress on the rapidity resummation for $B$-meson wave functions
which are key ingredients of TMD factorization formulae for radiative-leptonic, semi-leptonic and non-leptonic $B$-meson decays.
The crucial observation is that  rapidity resummation induces a strong  suppression of $B$-meson wave functions at small light-quark
momentum, strengthening the applicability of  TMD factorization in  exclusive $B$-meson decays. The phenomenological consequence
of rapidity-resummation  improved $B$-meson wave functions is further discussed in the context of $B \to \pi$ transition form factors
at large hadronic recoil.  }
\maketitle
\section{Introduction}
\label{intro}

QCD factorization theorems are indispensable to the theoretical descriptions  of high-energy processes probed at worldwide
collider experiments,  and  the  predictive power  lies on the process-independence  of non-perturbative hadronic functions,
entering the factorization formulae. $B$-meson light-cone distribution amplitudes (LCDAs) and transverse-momentum dependent (TMD)
wave functions are fundamental inputs for QCD calculations of exclusive $B$-meson decays applying collinear factorization
and TMD factorization theorems. Great efforts have been devoted to the studies of  renormalization  properties and perturbative
constraints of $B$-meson LCDAs in both momentum space \cite{Lange:2003pk,Braun:2003wx,Lee:2005gza,Bell:2008er} and  dual space \cite{Bell:2013tfa,Feldmann:2014ika}, and to the explorations of QCD evolution equations of $B$-meson TMD wave functions \cite{Li:2004ja,Li:2012md} with the standard renormalization-group (RG) method and the QCD resummation technique (for a recent
review, see \cite{Li:2013ela}).

Providing three-dimensional profile of a given hadron, TMD wave functions are more complicated due to the emergence of light-cone (or rapidity)
singularities in the end-point region, which cancel in the QCD corrections to LCDAs. Such rapidity divergences can be regularized by the
rapidity ($\zeta$) of non-light-cone Wilson lines, at the price of generating infrared enhanced double logarithm $\ln^2 \zeta^2$ \cite{Li:2012nk,Cheng:2014fwa} for $B$-meson wave functions and single logarithm $\ln \zeta^2$ \cite{Li:2010nn,Cheng:2014gba} 
for pion wave functions. Resummation of these rapidity logarithms in both hadronic wave functions and hard functions are essential 
to make scheme-independent predictions for physical observables. In the following, we will discuss the construction of rapidity evolution equation for $B$-meson TMD wave functions in the Mellin and impact parameter spaces in section \ref{evolution equation}, 
and present the  resummation improved $B$-meson wave functions in momentum  space by performing the inverse Mellin transformation in section  \ref{resummation of TMD} where the resummation effects on  $B \to \pi$ transition form factors are also reported.

\section{Rapidity Evolution Equation}
\label{evolution equation}

The TMD wave functions of $B$-meson are defined by the non-local vacuum-to-hadron matrix element in coordinate space
\cite{Li:2012md}
\begin{eqnarray}
& &\langle 0|{\bar q}(y) W_y(n)^{\dag}I_{n;y,0}W_0(n) \Gamma h(0)|{\bar B}(v)\rangle \nonumber \\
&& = -{i f_B m_B \over 4} {\rm Tr} \left \{ {1 +\slash \! \! \!   v \over 2}
\left [ 2 \, \Phi_{B}^{+}(t,y^2) + { \Phi_{B}^{-}(t,y^2)
-\Phi_{B}^{+}(t,y^2)   \over t }  \slash \! \! \!   y \right ] \gamma_5 \, \Gamma\right \},
\label{de1}
\end{eqnarray}
which depends on longitudinal ($t=v \cdot y)$ and transverse ($y^2$) variables  as well as the  non-light-like
gauge vector $n$ defining the Wilson lines.

It is demonstrated in \cite{Li:2012nk}  that  both the double rapidity logarithm  $\ln^2 \zeta^2$  with $\zeta^2=4(v \cdot n)^2/n^2$,
due to the overlap of the collinear enhancement from a loop momentum $l$  collimated to the gauge vector $n$ and the soft enhancement,
and the mixed logarithm $\ln \mu_f \, \ln \zeta^2$ ($\mu_f$ being the renormalization scale) which has the same origin
of cusp divergence in $B$-meson LCDAs \cite{Lange:2003pk} appear in the  next-to-leading-order (NLO) QCD corrections to $B$-meson wave functions. Simultaneous  resummation of two distinct types of rapidity logarithms for $B$-meson wavefunctions makes it subtler than that of the traditional Sudakov resummation for fast-moving light hadrons (see, however, \cite{Li:2013xna} for  recent progress).

Recalling that the Sudakov resummation of $\ln^2 \zeta^2$ for pion wave function can be achieved by constructing the
evolution equation from varying the gauge vector $n$, because the collinear divergence arises from the region with
a loop momentum collimated to the pion  momentum and varying the vector $n$ does not result in an additional collinear
divergence. This trick  however does not apply to the resummation of $B$-meson wave function due to the nature of rapidity logarithm
explained in the above. To resum the rapidity logarithm for $B$-meson wave functions, we utilize the fact that  the resulting
collinear divergence  is insensitive to the heavy-quark velocity $v$ and the effect from varying $v$ can then be factorized from
the TMD wave functions. Trading the rapidity derivative for the differential of velocity yields
\begin{equation}
\zeta^2 \frac{d}{d\zeta^2}\Phi_B^{(b)}(x,k_T,\zeta^2,\mu_f) =
\frac{v\cdot n}{2\epsilon_{\alpha\beta}v^\alpha n^\beta}v^+
\frac{d}{dv^+}\Phi_B^{(b)}(x,k_T,\zeta^2,\mu_f),\label{deri}
\end{equation}
where $\epsilon_{\alpha\beta}$ is an anti-symmetric tensor with $\epsilon_{+-}=-\epsilon_{-+}=1$,
$v^{+}$ is the plus component of the $b$-quark velocity and
$x$ is the longitudinal momentum fraction of the light quark.
Here,  the crucial point is that the velocity derivative will be only applied to  the Feynman rules
involving an effective heavy $b$-quark
\begin{equation}
\frac{v\cdot n}{2\epsilon_{\alpha\beta}v^\alpha
n^\beta}v^+\frac{d}{dv^+}\frac{v^\mu}{v\cdot l}=\frac{\hat
v^\mu}{v\cdot l},
\end{equation}
inducing  the special vertex
\begin{equation}
\hat v^\mu \equiv \frac{v\cdot n}{2\epsilon_{\alpha\beta}v^\alpha
n^\beta}\epsilon_{\rho \lambda} v^\rho \left(g^{\mu
\lambda}-\frac{v^\mu l^\lambda}{v\cdot l}\right).
\label{special vertex}
\end{equation}

\begin{figure}
\centering
\includegraphics[width=10 cm,clip]{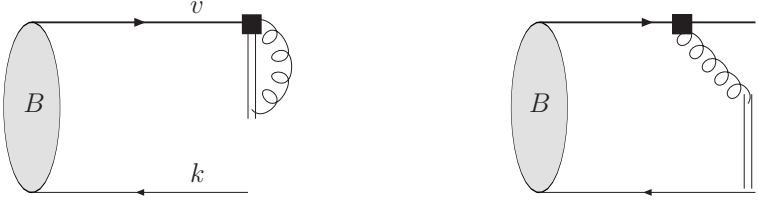}
\caption{Leading-order soft kernel for the  rapidity evolution equation, where the
square denotes the special vertex of Eq. (\ref{special vertex}).
Taken from \cite{Li:2012md}. }
\label{fig: kernel}       
\end{figure}

The rapidity evolution equation of $B$-meson wave function  can then be written as
\begin{eqnarray}
\zeta^2\frac{d}{d\zeta^2}\Phi_B(x,k_T,\zeta^2,\mu_f) &=&
K^{(b,1)}\otimes \Phi_B(x,k_T,\zeta^2,\mu_f) \nonumber \\
&& -
\frac{1}{Z_\Phi}\left(\zeta^2\frac{d}{d\zeta^2}Z_\Phi\right)\Phi_B(x,k_T,\zeta^2,\mu_f),
\label{RGE:momenrum}
\end{eqnarray}
where $Z_\Phi$ is the counterterm for the ultraviolet renormalization of TMD wave function.
The one-loop kernel $K^{(b,1)}$  collects the soft gluon dynamics as displayed in  figure \ref{fig: kernel}.

Computing  the two effective diagrams yields
\begin{eqnarray}
K_1^{(b,1)}&=& -{\alpha_s C_F \over 4 \pi} \, \Gamma(\epsilon) \left({
4 \pi \mu_f^2 \over \lambda^2}\right)^{\epsilon} \, \left( { v \cdot n
 \over \epsilon_{\alpha\beta}v^\alpha n^\beta  } \right)^2,\label{kb1}\\
\tilde K_2^{(1)}(N,b,\zeta^2)&=& {\alpha_s C_F \over 2 \pi} \, \left ( { v \cdot n  \over
\epsilon_{\alpha\beta}v^\alpha n^\beta  } \right )^2 \left [  K_0
\left(\lambda b \right) -  K_0\left(\sqrt{\zeta^2}
 \frac{m_B b}{N} \right) \right ], \label{delk}
\end{eqnarray}
where the gluon mass $\lambda$ is introduced to regularize the infrared divergence in each diagram,
Mellin and Fourier transformations for the $B$-meson wave function are performed  in the evaluation
of irreducible contribution illustrated in the second diagram. It is then straightforward to derive
\begin{equation}
\zeta^2\frac{d}{d\zeta^2}\tilde{\Phi}_B(N,b,\zeta^2,\mu_f)=
\tilde{K}^{(1)}(N,b,\zeta^2,\mu_f) \, \tilde{\Phi}_B(N,b,\zeta^2,\mu_f),
\label{RGE:renom scheme}
\end{equation}
with the renormalized evolution kernel
\begin{equation}
\tilde{K}^{(1)}(N,b,\zeta^2,\mu_f)= - {\alpha_s C_F \over 2 \pi} \,
\left [ \ln \frac{\mu_f b}{2} +\gamma_E +  K_0\left(\sqrt{\zeta^2}
\frac{m_B b}{N} \right) \right ],\label{K0}
\end{equation}
where the infrared divergence cancels exactly between the two effective diagrams.
The single logarithm of soft kernel $\tilde{K}^{(1)}$  can be organized by the RG resummation
\begin{equation}
{\cal K}^{(1)}(N,b,\zeta^2,\mu_f)=
\tilde{K}^{(1)}(N,b,\zeta^2,\mu_c)-\int^{\mu_f}_{\mu_c}
\frac{d\mu}{\mu}\lambda_K(\alpha_s(\mu)) \, \theta(\mu_f-\mu_c),
\label{improved scheme kernel}
\end{equation}
where the   scale  $\mu_c= a \, \sqrt{\zeta^2}\,m_B/N$, $a$ being an order-unity constant,
is adjusted to diminish the logarithmic enhancement in the initial condition.

In addition,  the factorization scale evolution of TMD wave function itself is computed as
\begin{equation}
\tilde{\Phi}_B(N,b,\zeta^2,\mu_f)= \exp \left [ -\int^{\mu_f}_{\mu_0} \frac{d\mu}{\mu}
\, {\alpha_s(\mu) \over 2 \pi}C_F \left ( \ln \zeta^2 -2   \right )    \right ] \,
\tilde{\Phi}_B(N,b,\zeta^2,\mu_0) \,. \label{scale solution}
\end{equation}
Combining the rapidity and scale evolutions and choosing $\mu_f = a \, \zeta_0  \, m_B$, we obtain
\begin{eqnarray}
\tilde{\Phi}_B(N,b)&=&\exp \left [
\int_{\zeta_0^2}^{N^2\zeta_0^2} { d \tilde{\zeta}^2 \over
\tilde{\zeta}^2 } {\cal K}^{(1)}(N,b,\tilde{\zeta}^2,\mu_f)
-\int^{\mu_f}_{\mu_0} \frac{d\mu}{\mu} \,
{\alpha_s(\mu) \over 2 \pi}C_F  \left ( \ln \zeta_0^2 -2 \right) \right] \nonumber \\
&& \times  \tilde{\Phi}_B(N,b,\zeta_0^2,\mu_0)\,, \label{solu}
\end{eqnarray}
with the simplified rapidity kernel
\begin{equation}
{\cal K}^{(1)}(N,b,\zeta^2,\mu_f)= -  {\alpha_s(\mu_c) \over 2
\pi} \, C_F \, \ln a
-\int^{\mu_f}_{\mu_c} \frac{d\mu}{\mu}  \, {\alpha_s(\mu) \over 2
\pi}C_F \, \theta(\mu_f-\mu_c)\, \label{k30}
\end{equation}
which is valid in the large $N$ limit.

\section{Resummation Improved $B$-meson TMD Wave Functions}
\label{resummation of TMD}

We are now in a position of discussing the resummation effect on $B$-meson wave functions in $x$-space
by performing the inverse Mellin transformation
\begin{equation}
{\Phi}_B^{\pm}(x,k_T)=\int_{c-i\infty}^{c+i\infty}\frac{dN}{2\pi i} (1-x)^{-N}
\tilde{\Phi}_B^{\pm}(N,k_T),
\end{equation}
for the solution given by  Eq. (\ref{solu}). To achieve this purpose, we need to know the initial  conditions of
TMD wave functions which are postulated to have the factorized form
\begin{equation}
\Phi_{B}^{\pm}(x,k_T,\zeta_0^2) =  \phi_{B}^{\pm}(x, \zeta_0^2) \,
\phi(k_T) \,,
\end{equation}
to reduce the numerical analysis.  The longitudinal parts are further taken from the so-called ``free-parton"  model
\cite{Kawamura:2001jm}
\begin{equation}
\phi_{B}^{+}(x, \zeta_0^2) = { x \over 2 x_0^2} \, \theta(2 x_0-x)  \qquad
\phi_{B}^{-}(x, \zeta_0^2)= { 2 x_0 -x \over 2 x_0^2} \, \theta(2 x_0-x), \, \label{phipi}
\end{equation}
which in Mellin space correspond to
\begin{equation}
\tilde{\phi}_{B}^{+}(N, k_T, \zeta_0^2) = {1- (1-2 x_0)^{N} (1+ 2 x_0 N) \over  2 x_0^2 N(N+1)}, \qquad
\tilde{\phi}_{B}^{-}(N, k_T, \zeta_0^2) =  { (1-2 x_0)^{N+1} + 2 x_0 N + 2 x_0-1
\over  2 x_0^2 N(N+1)}. \nonumber
\end{equation}

\begin{figure}[ht]
\begin{center}
\hspace{-1 cm}
\includegraphics[scale=0.5]{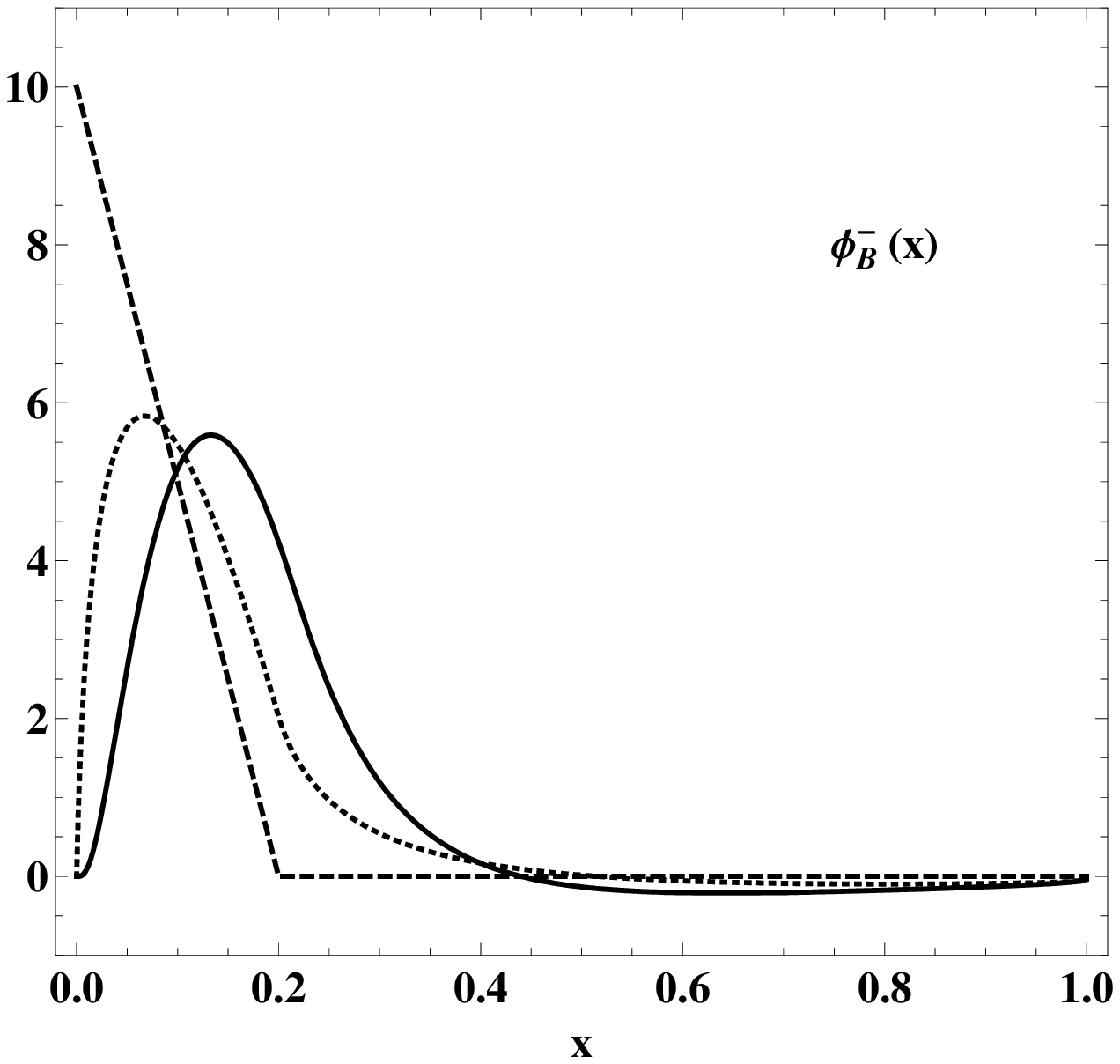}\hspace{0.5cm}
\includegraphics[scale=0.5]{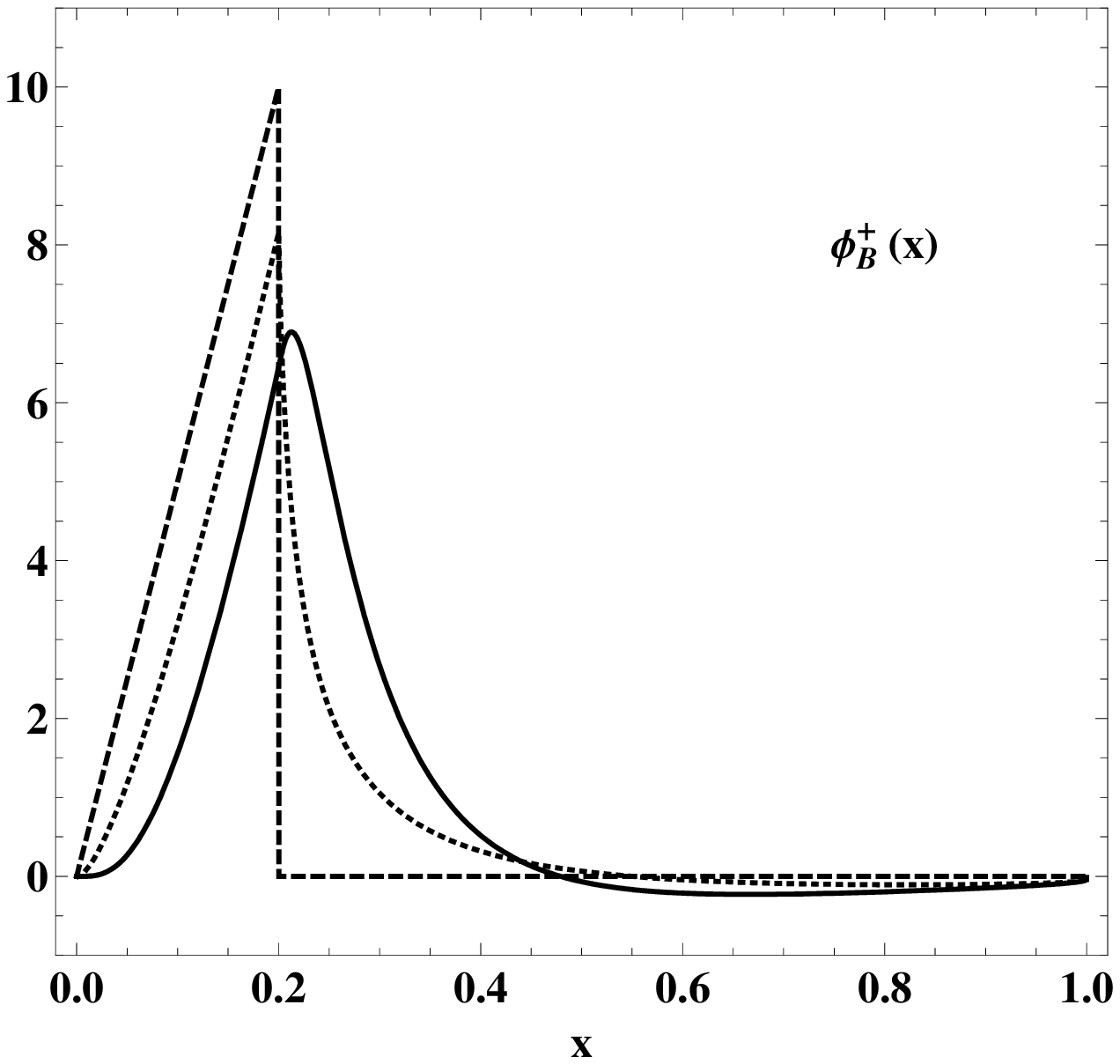}
\caption{The dashed, dotted and solid curves correspond to the $x$ dependence of initial condition
$\phi_{B}^{\pm}(x, \zeta_0^2)$, and the resummation improved $\phi_{B}^{\pm}(x)$ for fixed $\alpha_s=0.3$
and for running $\alpha_s$ with $\zeta_0=e/10$ and $a=1$. Taken from \cite{Li:2012md}. }
\label{fig: shape of phiBm}
\end{center}
\end{figure}

The inverse Mellin transformation will be firstly performed for the resummation improved wave functions
$\tilde{\Phi}_B^{\pm}(N,k_T)$ with frozen strong coupling constant $\alpha_s$, to make the analytical behavior
more transparent, and then with running $\alpha_s$.
The resummation effects on the $x$ dependence of $B$-meson wave functions $\phi_{B}^{\pm}(x)=\Phi_{B}^{\pm}(x, k_T)/\phi(k_T)$
are illustrated in figure \ref{fig: shape of phiBm}.
The primary observations are summarized as follows:
\begin{itemize}

\item  {Faster  than linear attenuation of $\phi^{\pm}(x)$  in the small $x$ region is realized after the  resummation improvement,
albeit with the non-vanishing $\phi^{-}(x)$ at $x=0$  initially. }

\item  {Resummation improved $B$-meson wave functions become smooth and develop radiative tails. }

\item  {The normalization conditions $\int_0^1 d x\, \phi_{B}^{\pm}(x)(x) =1$  are unaffected by the rapidity resummation.}

\end{itemize}

It will be interesting to inspect the rapidity resummation effect for phenomenological observables which
are relevant for the precision test of CKM mechanism (for recent reviews, see \cite{Bediaga:2012py,Wang:2014sba})
and for the hunting of beyond standard model physics. In this respect,  we consider the $B \to \pi$ transition form
factors $f_{B \pi}^{\pm}(q^2)$ at large hadronic recoil as illuminative examples, which are essential to the
golden-channel determination of matrix element $|V_{ub}|$ exclusively. Numerically the resummation effects are
found to decrease the above two form factors by approximately $25 \%$ at $q^2=0$, due to the strong suppression of
$B$-meson wave functions at the end-point. Notice that such improvement is sizeable and must be taken into account
in the future calculations of $B \to \pi$ form factors applying TMD factorization theorem to catch up with the
precision  achieved in the calculations from light-cone QCD sum rules \cite{Khodjamirian:2011ub,Bharucha:2012wy}.

\section{Conclusion}
\label{conclusion}

Applying the QCD resummation technique with non-light-like Wilson lines, we construct rapidity evolution equation
for TMD wave functions of $B$-meson collecting the  double logarithm  $\ln^2 \zeta^2$ and mixed  logarithm
$\ln \mu_f \, \ln \zeta^2$. Technically,  the rapidity resummation of $B$-meson wave functions is novel
due to  the different nature of collinear divergence in  QCD correction from that of energetic light mesons.
The resummation improved $B$-meson wave functions induce  a strong suppression in the
small $x$ region and hence enhance the applicability of TMD factorization in exclusive $B$-meson decays.
Sizeable corrections to the $B \to \pi$ transition form factors are also observed due to the resummation improvement.
Many open questions concerning the TMD wave functions of $B$-meson remain to be answered, and these include
(a) What are the relations between  TMD wave functions and LCDAs at large transverse momentum?
(b) What are the operator-product-expansion  constraints of TMD wave functions?
(c) Can we gain some insights of TMD wave functions at low energy scale from non-perturbative approaches?
As a final remark, the rapidity resummation technique presented here can be generalized immediately to the
TMD wave functions of  $\Lambda_b$- baryon entering the factorization formulae of  many exclusive  decays
\cite{Chou:2001bn,He:2006ud,Lu:2009cm} which are of increasing interest at the LHC and Tevatron.

\section*{Acknowledgements}

We are grateful to  Hsiang-nan Li for a very fruitful collaboration. 
YMW would like to thank the organizers of QCD@work 2014 for inviting him to this stimulating workshop
and for the generous finical support. This research is also partially supported by 
the National Science Foundation of China under Grant No. 11005100 (YLS), and by
the DFG Sonderforschungsbereich /Transregio 9 ``Computergest\"{u}tzte Theoretische Teilchenphysik" (YMW).

%
%
%

\end{document}